\documentclass[apl,epsfig,amsmath,amssymb,preprint]{revtex4}
\usepackage{graphicx}
\usepackage{dcolumn}
\usepackage{bm}

\begin{document}

\title{Reproducible Low Contact Resistance in Rubrene Single-Crystal Field-Effect Transistors with Nickel Source and Drain Electrodes}

\author{Iulian N. Hulea, Saverio Russo, Anna Molinari, and Alberto F.  Morpurgo }
\affiliation{Kavli Institute of Nanoscience, Delft University of
Technology, Lorentzweg 1, 2628CJ Delft, The Netherlands}

\date{\today}

\begin{abstract}
We have investigated the contact resistance of rubrene
single-crystal field-effect transistors (FETs) with Nickel
electrodes by performing scaling experiments on devices with
channel length ranging from 200 nm up to 300  $\mu$m. We find that
the contact resistance can be as low as 100 $\Omega$cm with
narrowly spread fluctuations. For comparison, we have also
performed scaling experiments on similar Gold-contacted devices,
and found that the reproducibility of FETs with Nickel electrodes
is largely superior. These results indicate that Nickel is a very
promising electrode material for the reproducible fabrication of
low resistance contacts in organic FETs.

\end{abstract}

\maketitle

The possibility to downscale organic field-effect transistors
(FETs) is currently hindered by the high contact resistance
present at the interface between the metal electrodes and the
organic semiconductor \cite{Burgi03}. One of the main experimental
problems in the study and optimization of the contact resistance
originates from the observed irreproducibility. In spite of the
large effort put in the investigation of contact effects
\cite{Burgi03,Malliaras,Pesavento04,Street02,Necliudov03,Meijer03},
the reason for both the high values and the irreproducibility of
the contact resistance are not currently understood. Many
different phenomena are likely to play an important role,
including the presence of grain boundaries at the metal/organic
interfaces, the interface fabrication process (e.g., metal
diffusion into the organic semiconductors and extrinsic damage
introduced during the device assembly process), fluctuations in
the work function of the metal electrodes, etc. Currently, the
problem seems to be particularly severe for oligomer-based
devices. Whereas for FETs based on a number of different polymers
it has been found that the contact resistance scales linearly with
the carrier mobility \cite{Hamadani04}, for transistors based on
oligomers a very broad range of contact resistance values have
been measured on identically prepared devices, and no systematic
behavior has been observed \cite{Meijer03}.

 To address the issue of contact resistance in oligomer transistors, we have
recently started the investigation of organic single-crystal FETs
with different metal contacts. Single-crystal devices are
particularly advantageous for this purpose because their
electrical characteristics exhibit an excellent level of
reproducibility from sample to sample \cite{deBoer03}. This is
crucial for a reliable comparison of FETs with different channel
length, i.e. to perform scaling experiments from which the value
of the contact resistance can be extracted.

In this paper we focus on rubrene single-crystal FETs with Nickel
electrodes. Nickel was chosen because, although it oxidizes in
air, its native oxide is conductive and has a work-function of 5.0
eV \cite{Oliveir01}, ideally suited to inject carriers into the
highest occupied molecular orbital of many molecular
semiconductors. By performing a conventional scaling analysis
\cite{Meijer03,Blanchet04,Hamadani04} of the electrical
characteristics of these devices we extract the value of the
contact resistance. We find values of $R_C$ as low as 100
$\Omega$cm, i.e. 50 times smaller than in the best oligomer FET
reported to date \cite{Meijer03}. The spread in values in the
contact resistance measured on transistors fabricated on the same
crystal is small (less than a factor of 2); devices fabricated on
different crystals exhibit a somewhat large spread, ranging from
100 $\Omega$cm to 1.5 k$\Omega$cm (and typically between 200
$\Omega$cm and 1 k$\Omega$cm), but still considerably smaller than
what has been observed so far in oligomer FETs. For comparison, we
have also investigated a number of single-crystal FETs contacted
with gold electrodes, the material commonly used for the
fabrication of contacts in organic transistors, and found a
considerably lower reproducibility level. This indicates that
Nickel is a very promising material for the fabrication of
contacts for organic transistors, even though the surface of the
electrodes oxidizes. We note that Nickel is also advantageous as
compared to gold because it is more mechanically robust, which
should minimize the possibility of (electro)migration into organic
materials during device operation, and cheaper.

\begin{figure}[t!]
\begin{center}
\includegraphics[width=8cm]{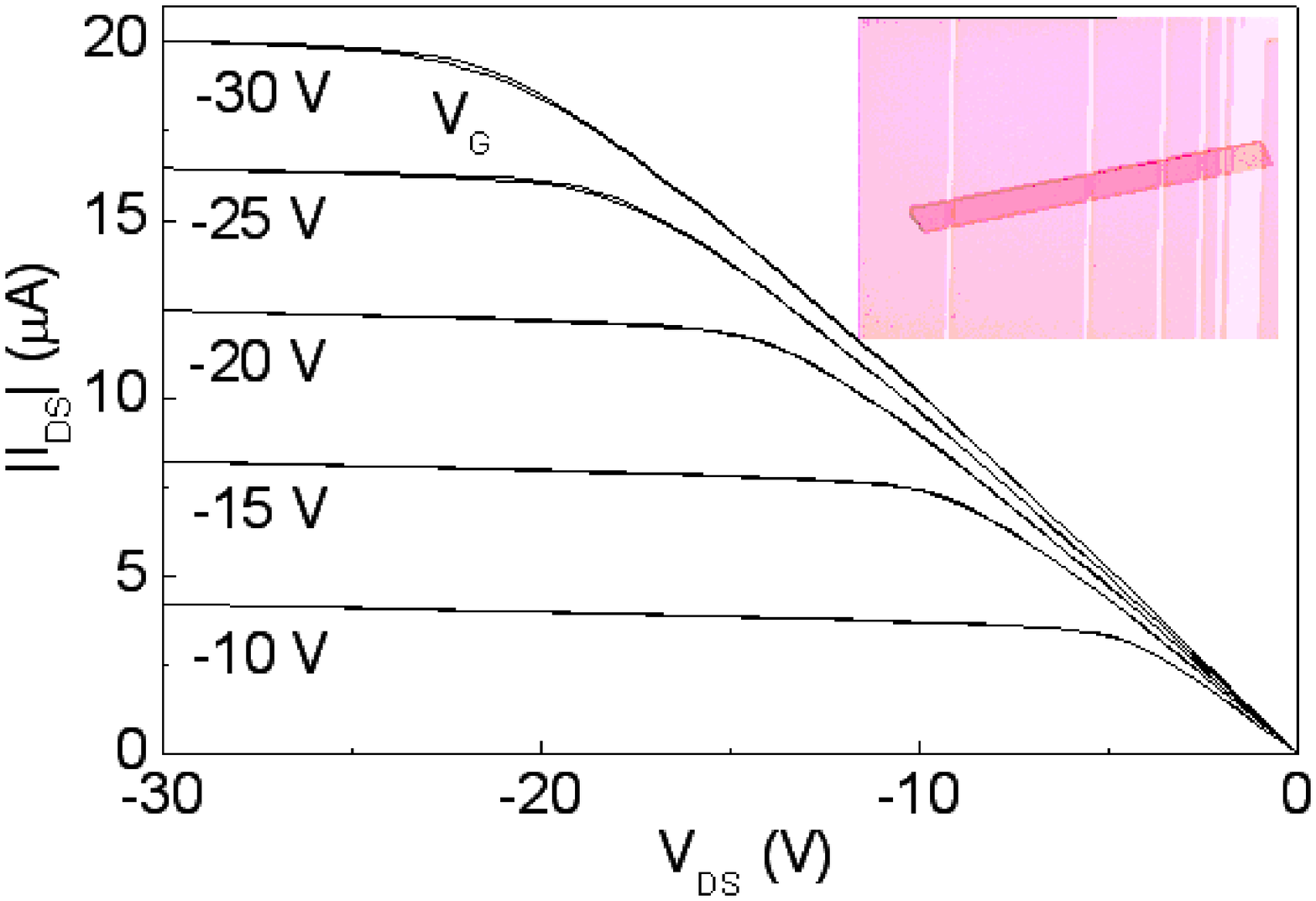}
\end{center}
\noindent{\caption{Typical transistor characteristics measured on
a rubrene single-crystal FET with Ni source-drain electrodes. The
inset shows a top view of one of the devices used in our
investigation (for this device the crystal width $W$ is 35
$\mu$m).} \label{Fig1}}
\end{figure}

The FET fabrication is based on electrostatic bonding of rubrene
single crystals to a doped silicon substrate (acting as a gate)
covered with a 200 nm- thick thermally grown SiO$_2$, with
prefabricated source and drain contacts (see Ref.\cite{deBoer04}
for details). The contacts are prepared by conventional optical or
e-beam lithography, nickel electron-beam evaporation (~20 nm), and
lift-off. The rubrene crystals are separately grown by means of a
vapor phase transport technique \cite{Laudise}; they are
millimeters long and their width and thickness are respectively of
the order of 100  $\mu$m and 1 $\mu$m. The device layout (see
Fig.\ref{Fig1}) is such that FETs with different channel lengths
are fabricated on the same single crystal. Many different samples
were studied with channel length ranging from 200 nm to 300
$\mu$m. Prior to the crystal adhesion, an oxygen plasma treatment
is performed to remove residues of resists possibly still present
on the SiO$_2$ surface. Although the exposure of the electrodes to
oxygen plasma contributes to the oxidation of the Nickel surface,
it does not preclude the realization of reproducible,
low-resistance electrodes.

 In the linear regime of transistor operation the total device resistance $R_T(L)$ can be
written as \cite{Meijer03,Blanchet04,Hamadani04}:
\begin{equation}\label{res}
R_T(L) = R_{ch}(L) + R_{C},
\end{equation}
 Here
\begin{equation}\label{res1}
R_{ch} = \frac{L}{W C_i(V_G-V_{TH})}\frac{1}{\mu}
\end{equation}
is the channel resistance and $R_C$ a length independent contact
resistance, $C_i$  is the capacitance of the insulating layer per
unit area, $V_G$ and $V_{TH}$ are the gate and the threshold
voltage, $W$ is the channel width, and $\mu$ the hole mobility.
The contact resistance is obtained by extrapolating the
experimental data to zero channel length. The slope of the
$R_T$-vs-$L$ curves also permits to extract the carrier mobility.
The comparison of the mobility value obtained from this slope with
the one obtained from the usual formula for the linear regime of
the individual FETs
\begin{equation}\label{mobility}
\mu = \frac{L}{W C_i V_{DS}}\frac{\partial I_{DS}}{\partial V_G}
\end{equation}
is used as a consistency check of our analysis.

 Fig.~\ref{Fig1} shows the electrical
characteristics measured on one of the FETs that we have
investigated and it is typical for all our Ni contacted devices.
The data have been measured with the FETs in high vacuum (p $<$
$10^{-6}$ mbar) and dark, using an Agilent E5270A or a HP 4192A
parameter analyzer. Usually, no hysteresis is observed in the
I$_{DS}$ -V$_{DS}$ plot at fixed gate voltage and in the I$_{DS}$
-V$_G$ at fixed source drain bias. The linearity of the I$_{DS}$
-V$_{DS}$ at low bias gives a first indication of a good contact
quality.

The scaling of the total device resistance $R_T$ versus device
length $L$ is shown in Fig. 2 for different values of the gate
voltage V$_G$, with V$_{DS}$ = -1 V, for a sample with channel
length in the range 20 -200  $\mu$m. Clearly, $R_T$ does scale
linearly with $L$, implying that for a given device the contact
resistance $R_C$ is approximately the same irrespective of the
channel length. The value of $R_C$ is then given by the intercept
at $L$=0. To compare the behavior of devices fabricated on
different crystals we normalize the contact resistance to the
channel width, i.e. we consider $R_C^* = R_C W$
\cite{Burgi03,Meijer03}. For all different samples (in total,
approximately 50 individual FETs were measured) we find values of
$R_C^*$  in between 100 $\Omega$cm and 1.5 k$\Omega$cm, and most
typically in between 200 $\Omega$cm and 1 k$\Omega$cm, at $V_G$=
-30 V, usually only very weakly dependent on gate voltage.

\begin{figure}[t!]
\begin{center}
\includegraphics[width=8cm]{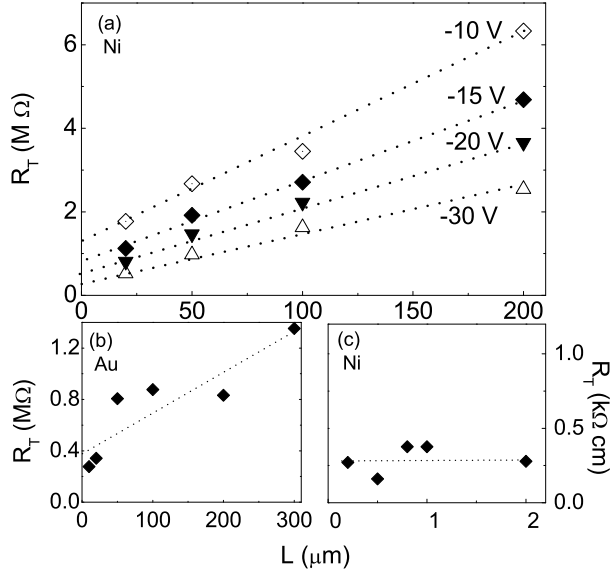}
\end{center}
\noindent{\caption{(a) Scaling of the device resistance for Nickel
contacted devices as a function of channel length for different
values of the gate voltages (W = 35 $\mu$m). The intercept at $L$
= 0 gives the contact resistance. (b) Similar scaling curve for a
gold-contacted FET: it is visible that the deviations from linear
scaling are larger in this devices as compared to Nickel-contacted
devices. In other gold-contacted FETs, the magnitude of the
fluctuations was larger than for the sample whose data are shown
here. (c) Normalized resistance measured on a Ni-contacted FETs
fabricated on the same rubrene crystal. $R_T$ does not depend on
$L$ because for $L <2 \mu$m the channel resistance is negligible
with respect to the contact resistance. In all panels, the lines
are a guide to the eye.} \label{fig2}}
\end{figure}

We have also analyzed the spread in contact resistance values for
FETs fabricated on the same crystal, by looking at devices with L
ranging from 200 nm to a few microns. Because the rubrene crystals have a  a
high mobility (2-6 cm$^2$/Vs), the contact resistance exceeds the channel resistance in devices whose channel length is less than approximately 5-10 $\mu$m. For these devices, the total resistance is essentially independent of channel length, as shown in Fig~\ref{fig2}c. These data also show that for FETs fabricated on the same crystal, the spread in contact resistance values is less than a factor of two. Thus, both for short and long channel devices, we conclude that the values of $R_C^*$ in Nickel-contacted Rubrene single crystal FETs are up to 50 times smaller than the smallest contact resistance (5
k$\Omega$cm) reported to date for oligomer-based FETs
\cite{Meijer03}, and that they exhibit a drastic improvement in
reproducibility as compared devices studied in the past.

\begin{figure}[t!]
\begin{center}
\includegraphics[width=6cm]{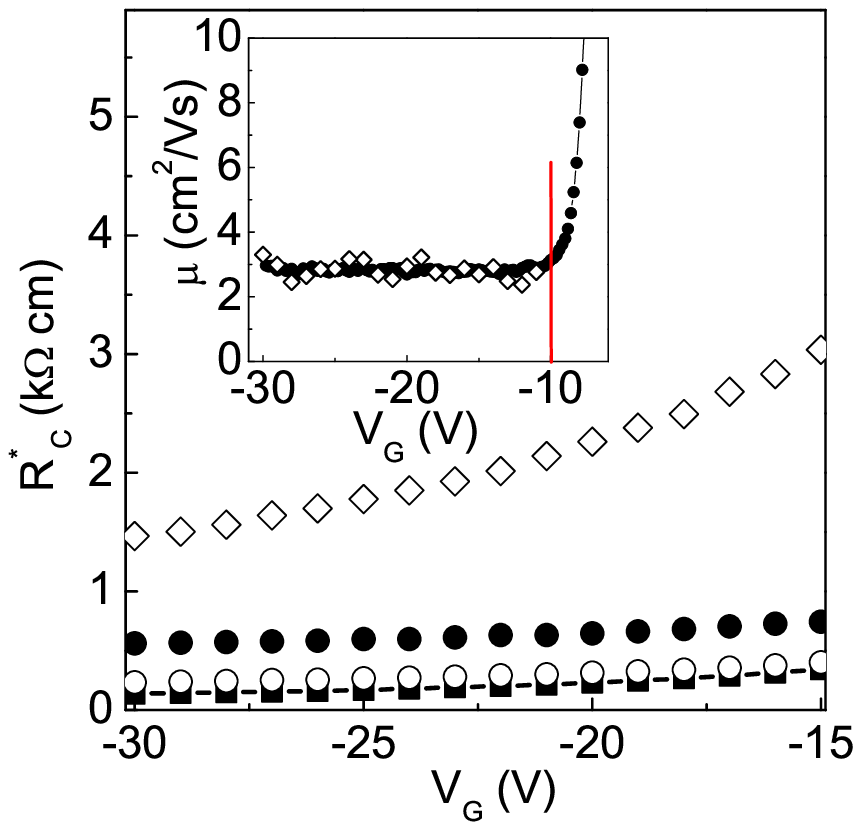}
\end{center}
\noindent{\caption{Gate voltage dependence of the normalized
contact resistance $R_C^*$ for four of the samples studied. The
insert shows the gate voltage dependence of the mobility $\mu$
determined from Eq. 3 (full circles) and from scaling $R_T(L)$
using Eq. 2 (open diamonds). The vertical line denotes the
beginning of the linear regime. For all the measurements V$_{DS}$=
-1 V} \label{fig3}}
\end{figure}

For devices with a channel length of 100 $\mu$m or longer, we have
calculated the value of mobility from the FET characteristics
using Eq. 3, as well as from the scaling analysis using Eq. 2. The
comparison of the values obtained in these two different ways (see
inset of Fig.\ref{fig3}) exhibits a remarkable agreement. This
agreement indicates the consistency of our analysis and gives full
confidence on the quantitative values obtained for the contact
resistance.

To understand if the low values and the reproducibility of the
contact resistance are due to the use of Nickel electrodes, or if
they are just a consequence of using high-quality organic single
crystals for the device fabrication, we have performed a scaling
analysis also for several gold-contacted single crystal FETs. In
all of these gold-contacted devices the mobility obtained via Eq.
3 (for long channel devices, L $>$ 100 $\mu$m) ranged from 2 to 6
cm$^2$/Vs, depending on the crystal \cite{Stassen04}. This
indicates that the crystal quality is the same for gold and
Nickel-contacted FETs. However, we found that in FETs with gold
electrodes the fluctuations in contact resistance are much larger
and in most cases prevent the observation of a clear scaling
between $R_T$  and $L$, for channel lengths comparable to or
smaller than 50-100 $\mu$m. This is illustrated in Fig~\ref{fig2}
b that shows the data for the gold-contacted devices which, among all devices measured, exhibited the best $R_T(L)$ scaling: as it is clear the fluctuations in measured resistance are much larger than for the Nickel-contacted devices (Fig~\ref{fig2}a and c). As a consequence of poor scaling, the data on gold-contacted transistors do not allow a precise determination of $R_C$ but only
a rough estimate of the lower limit, ~ 5 k$\Omega$cm, with a
spread of several times this value (even for FETs fabricated on
the same rubrene crystal).

From the above comparison, we conclude that Nickel does perform
better than gold as electrode material and that the
reproducibility in the contact resistance is not only due to the
use of single-crystalline material for the FET fabrication. As
gold has been used for contact fabrication in most of the organic
FETs fabricated in the past also in virtue of its chemical
inertness, the fact that Nickel does oxidize in air makes our
findings particularly unexpected. Nevertheless, the low contact
resistance values can be explained in terms of the work function
of oxidized Nickel that has been measured to be equal to 5.0 eV[6]
-ideal for hole injection into organic semiconductors- and by the
fact that non-stechiometric NiO$_x$ is a reasonably good
conductor. In this regard, it is also worth noting that recently
NiO$_x$ contacts have shown promising results as hole injectors in
organic light emitting diodes \cite{Chan04}.  Why oxidized Nickel
performs better than gold \cite{Lee05}, which is a better
conductor and has a comparable work function value, is less clear:
for its technological relevance, this issue deserves additional
investigations.

In conclusion, we have performed a scaling analysis of the
electrical characteristics of rubrene single-crystal FETs to show
that nickel can be used to fabricate source and drain electrodes
with an unprecedented low contact resistance and excellent
reproducibility.

This work was financially supported by FOM and is also part of the
NWO Vernieuwingsimpuls 2000 program.

\newpage

\newpage

\newpage
\begin{thebibliography}{10}

\bibitem{Burgi03}
{L. Burgi, T. J. Richards, R. H. Friend, and H. Sirringhaus,
Journal of Applied
  Physics {\bf 94}, 6129 (2003).}

\bibitem{Malliaras}
{ Y. L. Shen, A. R. Hosseini, M. H. Wong, G. G. Malliaras,
Chemphyschem {\bf
  5}, 16 (2004).}

\bibitem{Pesavento04}
{P. V. Pesavento, R. J. Chesterfield, C. R. Newman, and C. D.
Frisbie, Journal
  of Applied Physics {\bf 96}, 7312 (2004).}

\bibitem{Street02}
{R. A. Street and A. Salleo, Applied Physics Letters {\bf 81},
2887 (2002).}

\bibitem{Necliudov03}
{P. V. Necliudov, M. S. Shur, D. J. Grundlach, and T. N. Jackson,
Solid-State
  Electronics {\bf 47}, 259 (2003).}

\bibitem{Meijer03}
{E. J. Meijer, G. H. Gelinck, E. van Veenendaal, B. -H. Huisman,
D. M. de
  Leeuw, and T. Klapwijk, Applied Physics Letters {\bf 82}, 4576 (2003).}

\bibitem{Hamadani04}
{B. H. Hamadani and D. Natelson, Applied Physics Letters {\bf 84},
443 (2004).}

\bibitem{deBoer03}
{R. W. I. de Boer, M. E. Gerhenson, A. F. Morpurgo, and V.
Podzorov, Physica
  Status Solidi A- Applied Research {\bf 201}, 1302 (2004).}

\bibitem{Oliveir01}
{J. Olivier, B. Servet, M. Vergnolle, M. Mosca, and G. Garry,
Synthetic Metals
  {\bf 122}, 87 (2001).}

\bibitem{Blanchet04}
{G. B. Blanchet, C. R. Fincher, M. Lefenfeld, and J. A. Rogers,
Applied Physics
  Letters {\bf 84}, 296 (2004).}

\bibitem{deBoer04}
{R. W. I. de Boer, T. M. Klapwijk, and A. F. Morpurgo, Applied
Physics Letters
  {\bf 84}, 296 (2004).}

\bibitem{Laudise}
{R. A. Laudise, C. Kloc, P. G. Simpkins, and T. Siegrist, Journal
of Crystal
  Growth {\bf 187}, 449 (1998).}

\bibitem{Stassen04}
{A. F. Stassen, R. W. I. de Boer, N. N. Iosad, and A. F. Morpurgo,
Applied
  Physics Letters {\bf 85}, 3899 (2004).}

\bibitem{Chan04}
{I. M. Chan and C. Hong, Thin Solid Films {\bf 450}, 204 (2004).}

\bibitem{Lee05}
{Note that very recently pentacene thin-film FETs with evaporated
NiOx contacts
  have been claimed to have better performance than identically prepared
  devices with gold contacts. See J. Lee, D. K. Hwang, J. M. Choi, K. Lee, J.
  H. Kim, S. Im, J. H. Park and E. Kim, Applied Physics Letters {\bf 87},
  023504 (2005). In this work, however, the value of the contact resistance was
  not measured.}

\end{thebibliography}
\end{document}